\documentclass[prl,twocolumn,showpacs,preprintnumbers,amsmath,amssymb]{revtex4}
\usepackage{amsfonts}
\usepackage{graphicx}
\usepackage{dcolumn}
\usepackage{float}
\usepackage{bm}

\newcommand{\ind}[1]{\textrm{#1}}
\newcommand{\av}[1]{\langle {#1} \rangle}
\newcommand{\bra}[1]{\langle {#1} |}
\newcommand{\ket}[1]{| {#1} \rangle}
\renewcommand{\O}{ {\cal{O}} }

\newcommand{\com}[1]{}
\newcommand{\J}{\mathcal{J}}
\newcommand{\q}{\mathbf{q}}

\begin{document}

\title{Macroscopic two-state systems in trapped atomic condensates}

\author{Dmitry Solenov\footnote{E-mail: solenov@lanl.gov} and Dmitry Mozyrsky\footnote{E-mail: mozyrsky@lanl.gov}}

\affiliation{Theoretical Division (T-4) and the Center for
Nonlinear Studies (CNLS), Los Alamos National Laboratory, Los
Alamos, NM 87545, USA}

\date{\today}

\begin{abstract}

We consider a macroscopic two-sate system based on
persistent current states of a Bose-Einstein condensate (BEC) of interacting
neutral atoms confined in a ring with a weak Josephson link. We
demonstrate that macroscopic superpositions of different BEC flows
are energetically favorable in this system. Moreover, a
macroscopic two-state dynamics emerges in the low energy limit. We
also investigate fundamental limitations due to the noise inherent
to the interacting BEC of Josephson-ring geometry. We show that
the coherent macroscopic dynamics is readily measurable for an
experimentally accessible range of parameters.
\end{abstract}


\maketitle

Realization of macroscopic quantum two-state systems has been a challenge for
cold-atom BEC physics from early stages of its experimental
development. The research in this area has been primarily focused
on many-particle dynamics within double-well trapping potential
\cite{AndrewsKetterle, DWJosephson}. At the same time, despite an
appealing similarity with the microscopic single particle
two-state system, this geometry provides no easy way to achieve a
superposition of distinct many-body quantum states. Indeed, a
collection of non-interacting (or weakly interacting) boson atoms
confined in a double-well trap condenses into a BEC with the
``product" wave function not suitable to form a macroscopic
two-sate configuration. Repulsively interacting particles in such
trap favor \cite{LeggettBEC} Fock states---the system enters Mott
or ``Coulomb blockade"-like regime. Hence, attractively
interacting particles become the only option in such geometry. In
the latter case, atoms correlate \cite{Carr} forming a
``Schrodinger cat" state $\Psi^{DW}_{\lambda<0} = [\prod_i
\psi_L(\mathbf{r}_i) + \prod_i \psi_R(\mathbf{r}_i)]/\sqrt{2}$,
where $\psi_{L/R}(\mathbf{r})$ denote the single particle states
localized the left/right well of the trapping potential.
Observation of this macroscopic superposition, however, is
extremely challenging. The only manifestation of a coherent
superposition is the presence of the off-diagonal matrix elements
in the macroscopic two-state basis. In the case of the double-well
trapping potential these matrix elements are proportional to the
probability of all $N$ particles to tunnel through the barrier, which is
extremely small.

In this paper we study a cold atom based macroscopic two-state
system (a qubit) based on a persistent current BEC-Josephson
system \cite{TorBEC}. The two macroscopic states are metastable
current-carrying states of a BEC confined in a Josephson ring trap
\cite{BuchlerDidier,SolenovMozyrskyRING}. Such systems have become
experimentally available due to recent successes in dynamical BEC
trapping \cite{Boshier}. We start by deriving an effective
Schrodinger equation describing the low-lying energy states of the
system in terms of the phase difference across the Josephson
junction. For that we introduce a simple anzats that parameterizes
the low-lying states as a superposition of the condensate states
with different phases. Then we consider a better ground state
anzats that accounts for the fluctuations in the systems arising
from the interparticle interactions. We briefly discuss their
influence on system's dynamics and then analyze their effect on
the detection of the systems's state. Particularly, we show that
the admixture of "non-condensate" particles acts as an effective
noise in the time-of-flight (TOF) images of the BEC and therefore
imposes limitations on the resolution of such measurements. We
briefly discuss these fundamental limitations and argue that
weakly coupled Bose systems provide a good candidate to observe
macroscopic quantum tunneling (MQT) and macroscopic quantum
superpositions.

We consider a system
of $N$ locally interacting bosons confined to the rotating external potential $V$ corresponding
to a ring with a thin cut (a barrier). We assume that the transverse dimension of the ring
is small compared to the healing length of the bosons and therefore the system
can be considered effectively one-dimensional. The Hamiltonian of such system can be written as
\cite{LeggettBEC}
\begin{equation}\label{eq:H}
\hat H\!\! =\!\!\!
\sum_{n=1}^{N}\!\left[\!\frac{(i\nabla\!_n\!)^2}{2m}\! +\!
\Omega R\cdot i\nabla\!_n \!+\!\!
V\!(r_n)\!\right]\!\! +\!
\frac{\lambda}{2}\!\!\sum_{n\neq
m}^{N}\!\!\delta(r_n-r_m),
\end{equation}
where $\Omega $ rotation frequency of the ring (i.e., the barrier), $R$ is the radius of the ring,
$\lambda = 4\pi a/S m$, where $a$ is scattering length and $S$ is the ring's cross section area. Here and in the following we will use units with $\hbar=1$.

We start by evaluating the energy of the ground state of the
Hamiltoian (\ref{eq:H}) using Gross-Pitaevskii approach. That is,
we assume that the ground state wavefunction is a product,
$\Psi(r_1,...,r_N)=\chi(r_1)...\chi(r_N)$ and minimize the the
functional $E=\int dr_1...dr_N \Psi^\ast {\hat H}\Psi$ by varying
it with respect to the single-particle state $\chi$. It is
obvious, however, that, since the system is homogenous everywhere
except in the small region at the barrier, $|\chi(r)|^2\simeq {\rm
const}$ or $\chi\sim e^{i\Phi(r)}$ (except in the vicinity of the
barrier). Moreover, the phase-dependent terms in $E$ are $\sim
\int dr [(\nabla\Phi)^2/(2m)+\Omega R\nabla\Phi]$, and therefore
$\Phi$ must be a linear function of distance along the ring, i.
e., $\chi\sim e^{i\phi\theta/2\pi}$, where $\phi$ is the phase
difference across the barrier and $\theta$ is the azimuthal angle
parameterizing position along the ring. Evaluation of $E$ for such
product state yields $N\phi^2/2mL^2 -N\Omega\phi/2\pi+\lambda
N(N-1)/2L$, where $L$ is the circumference of the ring. This
expression, obviously, does not account for the contribution due
to the barrier, i.e., the Josephson energy. At the barrier the
particle density, $N|\chi(r)|^2$ is strongly dependent on $r$ and
therefore must be calculated self-consistently. It can be shown,
however, that to a good approximation, contribution of the barrier
region into the system's energy can be cast in the form $-E_\J
\cos\phi$, where the Josephson energy $E_\J$ is independent on
$\phi$ \cite{Joseph}. Thus we find that
\begin{equation}\label{eq:E(phi)}
E(\phi) \!= \!{\lambda N(N\!\!-\!\!1)}/{2L} +\!
N(\phi-\!\phi_0)^2\!/2mL^2 \!-\! E_\mathcal{J}\cos\phi,
\end{equation}
where $\phi_0 = mL^2\Omega/2\pi$.

When $\phi_0=\pi$ the effective potential
is a symmetric double well, corresponding to two macroscopically different states,
i. e., carrying different persistent currents. One can see, however, that such degeneracy
is lifted by quantum fluctuations, leading to macroscopic quantum tunneling. In order to see
this, let us evaluate the system's energy for the superposition state $C \int d\phi \Psi^{(0)}_\phi$,
where $\Psi^{(0)}_\phi$ is the above product state and $C$ is normalization constant. A straightforward
calculation shows that the expectation value of the interaction energy, i.e., of the last term in Eq.~(\ref{eq:H}),
is lower by $\lambda N/2L$ than that for the localized (in $\phi$) state $\Psi^{(0)}_\phi$.

We note, however, that while the interaction part of the
Hamiltonian in Eq.~(\ref{eq:H}) favors superposition, the first
two terms , obviously, ``prefer'' the localized state. Therefore
we search an optimal ground state wavefunction in the following
form~\cite{LeggettPsi, LeggettFrance}:
\begin{equation}\label{eq:PsiAll}
\Psi(\mathbf{r}_1,...,\mathbf{r}_N) = \sqrt{N/24\pi}\int
d\phi\psi(\phi)\Psi^{(0)}_\phi(\mathbf{r}_1,...,\mathbf{r}_N),
\end{equation}
where $\psi(\phi)$ is to be defined by minimization. In order to evaluate expectation value of
the energy for the wavefunction in Eq.~(\ref{eq:PsiAll}) we note that
the states $\Psi^{(0)}_\phi$ are approximately orthogonal. Indeed,
$\delta_N(\phi-\phi')\equiv\int d\mathbf{r}_1...d\mathbf{r}_N
\Psi^{(0)*}_\phi\Psi^{(0)}_{\phi'}$ yields
$\delta_N(\phi)=[\sin(\phi/2)/{\phi/2}]^N\thickapprox
e^{-N\phi^2/24}$. This is a rapidly varying function---the
combination $\sqrt{N/24\pi}\delta_N(\phi)$ approaches the true
$\delta$-function for $N\gg 1$. Moreover, one can see that for any
few-particle operator $\hat A$ the following identity holds:
\begin{equation}\label{eq:rule}
\bra{\Psi^{(0)}_\phi}{\hat A}\ket{\Psi^{(0)}_{\phi'}} =
\delta_N(\phi-\phi')\bra{\Psi^{(0)}_\phi}{\hat
A}\ket{\Psi^{(0)}_\phi}[1\!+\!\O(1/N)].
\end{equation}
As the result, one could think
that the expectation value of (\ref{eq:H}) calculated with respect to the
wavefunction (\ref{eq:PsiAll}) is merely
$\int d\phi |\psi(\phi)|^2 \bra{\Psi^{(0)}_\phi}{\hat H}\ket{\Psi^{(0)}_{\phi}} =
\int d\phi |\psi(\phi)|^2 E(\phi)$. This, however, is not the case. The
expectation value of the interaction part of the Hamiltonian is amplified by $N^2$ and the
neglected $\O(1/N)$ terms become important---the terms appearing
due to the finite width of $\delta_N(\phi)$ contribute to the
first order in $N$. A more careful evaluation of the interaction
expectation value gives
\begin{equation}\label{eq:E_lambda}
\frac{\lambda N\!(N\!\!-\!1\!)\!\!}{2L}\!\!\int\!\!\! d\phi d\phi
' \psi(\phi'\!)^*
\psi(\phi)\sqrt{\!\frac{N}{\!24\pi}}\delta_N\!(\phi'\!-\!\phi)
\frac{(\phi'\!\!-\!\!\phi)/2}{\tan\frac{\phi'-\phi}{2}}.
\end{equation}
The finite width of $\delta_N(\phi'-\phi)$ is clearly
non-negligible. We use the identity $N^2\phi^2\delta_N(\phi)/12 =
12\partial_{\phi}^2\delta_N(\phi) + N\delta_N(\phi)$ and evaluate
the integral over $\phi$ containing
$\partial_{\phi}\partial_{\phi}\delta_N(\phi'-\phi)$ by parts,
finally arriving at the result correct to $\O(N)$
\begin{equation}\label{eq:E-final}
E\!\!=\!\!-\frac{\lambda N}{2L} +\!\! \int\!\!\!
d\phi
\psi^*\!(\phi)\!\!\left[-\frac{6\lambda}{L}\partial_\phi^2\! +
\!E(\phi)\right]\!\!\psi(\phi ).
\end{equation}
The first term in the right hand side  of Eq.~(\ref{eq:E-final}) is the negative offset
mentioned earlier. The second term is a positive gain in energy due to variation of $\phi$.
Together with the last term in Eq.~(\ref{eq:E-final}) it can be viewed as kinetic and potential
energies of a ``phase-particle'', whose dynamics obeys Schrodinger equation $\hat H_{\rm eff}\psi =E\psi$,
where $\hat H_{\rm eff}= -(6\lambda/L)\partial^2_\phi +E(\phi)$.

The ground state for such ``particle'' is the symmetric
superposition of two states in each well of the effective
potential $E(\phi)$, separated from the next excited state, i.e.,
the antisymmetric combination, by the tunnel splitting
\begin{equation}\label{eq:gap}
\Delta E \sim {N\over m L^2\eta}\exp[-\eta\sqrt{\alpha}\delta\phi^3/72],\quad\quad
\delta\phi=\phi_R-\phi_L,
\end{equation}
where the exponential prefactor is the frequency of small
oscillation in each well, $\phi_L$ and $\phi_R$ are positions of
each minima, $\eta = \sqrt{NS/4\pi aL}$ is Tonks parameter
\cite{Tonks} (typically $\eta = 10-100$ for cold atom systems),
and $\alpha = mE_\J L^2/N$. When potential barrier is small,
$\alpha\simeq 1$ \cite{com1}. Energy $\Delta E$ defines the
timescale of tunneling transitions between states with two
different persistent currents.

The states with different persistent currents can be readily
detected by looking at absorbtion images of the density
distributions in time-of-flight (TOF) measurements \cite{Boshier,
SolenovMozyrskyRING}, e.g. Fig. (1). For long TOF $t_0$ single
particle wavefunctions $\chi_\phi^{TOF}(\mathbf{r})$ become
Fourier transforms of the initial single particle wavefunctions,
$\chi_\phi^{TOF}(\mathbf{r}) \sim
\chi_\phi(\mathbf{q})|_{\mathbf{q}= 2m\mathbf{r}/t_0}$. The
outcome of the measurement is best understood by considering
different moments of $\rho(\mathbf{q})$. A straightforward
calculations yields
\begin{eqnarray}\label{eq:moments}
\av{\hat\rho(\q)}&=&\int d\phi |\psi(\phi)|^2\rho_\phi(\q)
\\\nonumber
\av{\hat\rho(\q)\hat\rho(\q')}&=&\int d\phi
|\psi(\phi)|^2\rho_\phi(\q)\rho_\phi(\q')
\\\nonumber
&...&
\end{eqnarray}
where $\rho_\phi(\q)=(N/S)|\chi_\phi(\q)|^2$. These moments correspond to
a stochastic process: the probability to observe the outcome
$\rho_\phi(2m\mathbf{r}/t_\ind{TOF})$ and, hence, a particular
value of $\phi$, is given by $|\psi(\phi)|^2$. Therefore, the
measurement ``chooses" a single term in the sum (\ref{eq:PsiAll}),
and the function $\psi(\phi)$ becomes a quantum-mechanical wave
function of a macroscopic object---the superfluid current of the
entire condensate.
\begin{figure}
\includegraphics[width=7.0cm]{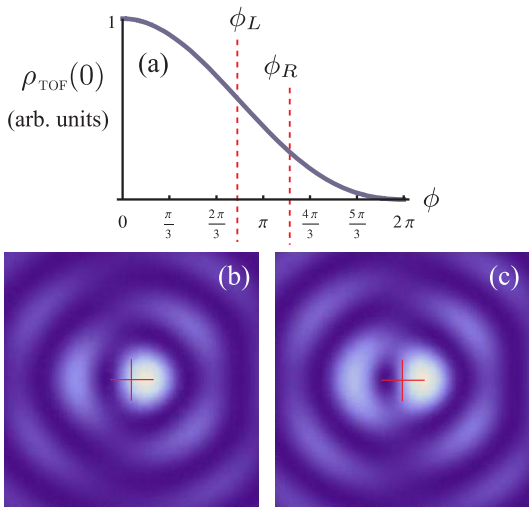}
\caption{(Color online) Expected TOF images of the Josephson-ring
system in a double-well regime. (a) TOF density at the center as a
function of $\phi$. Typical double-well potential minima $\phi_L$,
$\phi_R$ are shown to give relevant scale on $\phi$. The typical
TOF image corresponding to measurement outcome (b) $\phi_L$ and
(c) $\phi_R$. In both cases the cross shows the center of the
image.}\label{TOF-DW.eps}
\end{figure}

The above picture obviously is an approximation: We have assumed that all
particles are ``in the condensate'', e.g., Eq.~(\ref{eq:PsiAll}). For weakly interacting bosons the ground state is no longer a product,
but the well-known Bogolubov state. Therefore we consider a modified ground state wavefunction, still
in the form of Eq.~(\ref{eq:PsiAll}), but with modified $\Psi^{(0)}_\phi$, i.e.,
\begin{equation}\label{eq:Psi-aa}
\ket{\Psi^{(0)}_\phi}\rightarrow\ket{\Psi_\phi} \sim (\sum_n c_n a^\dag_{\phi+2\pi n}a^\dag_{\phi-2\pi
n})^{N/2}\ket{0},
\end{equation}
where $a^{(\dag)}_{\phi+2\pi n}$ is a creation (annihilation)
operator corresponding to a single-particle state
$\chi_{\phi,n}(\mathbf{r})=\chi_\phi(\mathbf{r})e^{in\theta}$, and
$c_n$ are variational coefficients \cite{com2}. By choosing  the
trial wavefunction in the form of Eq.~(\ref{eq:Psi-aa}) we assume
that the dynamics is adiabatic: The slow collective variable
$\phi$ is coupled to fast degrees of freedom, i.e., the
fluctuations. Indeed, due to the finite size of the the system the
quasiparticle spectrum is discrete with gaps $\sim ({\rm speed\
of\ sound}/L)= N/(m L^2\eta)$, which is much greater than the
energy scale associated with the tunneling between two wells,
e.g.,  Eq.~(\ref{eq:gap}). As a result the quasiparticles remain
in the ground state and readjust to the variations in $\phi$.

Corrections to the effective energy functional, e.g.,
Eq.~(\ref{eq:E-final}), turn out to be small in the limit of weak
interactions. The new ground state energy can be evaluated by
using Eq.~(\ref{eq:rule}), which, as can be directly verified,
still holds for the modified wavefunction, e.g.,
Eq.~(\ref{eq:Psi-aa}). Then, evaluating $\langle \Psi_\phi|{\hat
H}|\Psi_\phi\rangle$ and varying $c_n$'s (see Ref.
\cite{LeggettBEC} for similar calculation) we find that
corrections to the functional $E$ in Eq.~(\ref{eq:E-final}) are
suppressed by factor $1/\eta$, .i.e., are small in the weak
interaction limit \cite{com3}.

The effect of fluctuations, however, turns out to be quite
appreciable as far as measurement of close current-carrying states
is concerned. Indeed, while the number of non-condensate particles
is small, the difference between density profiles  (in the TOF
measurements) corresponding to different values of $\phi$, i.e.,
different persistent currents, is also small. In the opposite case
the barrier is to high to allow for tunneling on a reasonable time
scale. Moreover, as we will see, both the tunneling exponent in
Eq.~(\ref{eq:gap}) and the signal-to-noise ratio due to
"non-condensate" particles (to be defined below) are controlled by
the same parameters. To see this let us evaluate the set of
correlation functions as in Eq.~(\ref{eq:moments}), but for the
modified wavefunction given by Eq.~(\ref{eq:Psi-aa}). In doing so
we again use Eq.~(\ref{eq:rule}), and so the calculation reduces
to the evaluation of averages $\langle
\Psi_\phi|{\hat\rho}(q){\hat\rho}(q^\prime)...|\Psi_\phi\rangle$,
where TOF density operator ${\hat\rho}(q)=\sum_{n,n^\prime}
\chi^\ast_{\phi,n}(q)\chi_{\phi,n^\prime}(q)a^\dag_n a_{n^\prime}$
(again $\chi_{\phi,n}(q)$ is the Fourier transform of
$\chi_{\phi,n}(r)$). The first two moments are
\begin{eqnarray}\label{eq:new-moments}
\av{\hat\rho(\q)}&=&\int d\phi |\psi(\phi)|^2\tilde\rho_\phi(\q)
\\\nonumber
\av{\hat\rho(\q)\hat\rho(\q')}&=&\int d\phi
|\psi(\phi)|^2\left[\tilde\rho_\phi(\q)\tilde\rho_\phi(\q') +
\Lambda_\phi(\q,\q') \right]
\\\nonumber
\end{eqnarray}
where $\tilde\rho_\phi(\q) =\rho_\phi(\q) + \sum_{n\neq
0} \langle a^\dag_{\phi,n} a_{\phi,n}\rangle |\chi_{\phi,n}(\q)|^2$ is the renormalized density and
\begin{eqnarray}\label{eq:lambda}
\Lambda_\phi(\q,\q') = N \sum_n [\chi^\ast_{\phi,n}(\q)\chi_{\phi,n}(\q')\langle a^\dag_{\phi+2\pi n} a_{\phi+2\pi n}\rangle
\\\nonumber
+ \chi_{\phi,n}(\q)\chi_{\phi,n}(\q')\langle a_{\phi+2\pi n} a_{\phi+2\pi n}\rangle + {\rm c.\ c.}]
.
\end{eqnarray}
In evaluating Eqs.~(\ref{eq:new-moments}, \ref{eq:lambda}) we have
used Wick's theorem as well as the fact that in the limit of weak
interactions $\langle a^\dag_{\phi+2\pi n} a_{\phi+2\pi n}\rangle
|_{n=0}\simeq N\gg \langle a^\dag_{\phi+2\pi n} a_{\phi+2\pi
n}\rangle |_{n\neq 0}$ and therefore products $\langle
a^\dag_{\phi+2\pi n} a_{\phi+2\pi n}\rangle \langle
a^\dag_{\phi+2\pi n'} a_{\phi+2\pi n'}\rangle$, etc., with $n,\
n'\neq 0$, can be neglected. Moreover, since
$\Lambda_\phi(\q,\q')$ is small compared to
$\tilde\rho_\phi(\q)\tilde\rho_\phi(\q')$, we can replace it by
$\Lambda_\pi(\q,\q')$ in Eq.~(\ref{eq:new-moments}), - we are
interested in the situation when the barrier separating the two
persistent currents is small, $\delta\phi\ll \pi$. As a result
$\Lambda_\pi(\q,\q')$ can be viewed as the correlation function of
noise superimposed with ``signal''  $\tilde\rho_\phi(\q)$. It is
natural, therefore, to introduce a signal to noise ratio (SNR) as
\begin{equation}\label{eq:SNR}
\mathrm{SNR} = \frac{[\rho_{\pi-\delta\phi/2(0)}
-\rho_{\pi+\delta\phi/2}(0)]^2}{\Lambda_\pi(0,0)}
\end{equation}
where we define the strength of the signal as the difference in
the TOF particle densities at the center for two realizations with
phases $\pi\pm\delta\phi$, e.g., Fig. 1. The SNR in
Eq.~(\ref{eq:SNR}) can be easily evaluated: At the center
$\chi_{\phi,n}(\q=0) = B/(\phi/2\pi+n)$, where $B=(1/2\pi)\int
d{\bold r}\chi_\pi({\bold r})$. Then we obtain that
\begin{equation}\label{eq:F2}
\Lambda_\pi\!(0,\!0)\!= {4N^2B^4\over \pi^4} \!\sum_{n\neq
0}\!\!\left[\!\frac{\av{a^\dag_na_n}}{1\!\!-\!4n^2\!} \!+\!
\frac{\av{a_na_{-n}}}{(1\!\!+\!\!2n)^2\!}+c.c.
\right],
\end{equation}
where a shortcut $a_n^{(\dag)}=a_{\pi+2\pi n}^{(\dag)}$ has been
used to reduce notations. The averages in Eqs.~(\ref{eq:F2}) can
be found from the normal and anomalous Green's functions of the
interacting boson system, i.e. $\av{a_n^\dag a_n} = \int
\frac{d\omega}{2\pi}\mathcal{G}(n,i\omega)$ and $\av{a_n a_{-n}} =
\int \frac{d\omega}{2\pi}\mathcal{F}(n,i\omega)$, where
$\mathcal{G}(n,i\omega)=\frac{-i\omega -
\epsilon^\phi_n}{\omega^2+(\varepsilon^\phi_n)^2}$ and
$\mathcal{F}(n,i\omega)=\frac{{\lambda\rho}/{2}}{\omega^2+(\varepsilon^\phi_n)^2}$,
with $\epsilon^\phi_n={\hbar^2(\phi-\phi_0+2\pi n)}/{2mL^2}+
{\lambda\rho}/{2}$ and $\varepsilon^\phi_n = \sqrt{
(\epsilon^\phi_n)^2 - ({\lambda\rho}/{2})^2}$ \cite{AGD}. After
straightforward calculation we find that
\begin{equation}\label{eq:SNR2}
\mathrm{SNR} \simeq 7.84\eta\delta\phi^2.
\end{equation}

Eq.~(\ref{eq:SNR2}) is the principal result of this paper. It
reflects the fundamental difference between the phase dynamics of
the Josephson qubits and single body quantum mechanics: While
quantum mechanics, in principle, allows one to measure particle's
coordinate with infinite precision, the accuracy of the phase
measurements in Josephson qubits is limited by their many-body
nature. Remarkably, the precision for such measurement is
controlled by a single parameter $\eta$, i.e., the interaction
strength. Moreover, comparing Eq.~(\ref{eq:gap}) with
Eq.~(\ref{eq:SNR2}) we see that both the tunneling exponent and
the SNR are controlled by the same parameters. This fact is not
surprising: while stronger interactions enhance MQT, e.g.,
Eq.~(\ref{eq:gap}), they decrease the SNR due to the suppression
of the condensate density for stronger scattering. The tunneling
exponent, however, increases faster (in absolute value) with the
growth of $\delta\phi$ and therefore to keep it small together
with the condition $\mathrm{SNR}\gg 1$ we need $\eta\gg 1$. Thus
coherent MQT is observable only in weakly interacting BEC systems.
For typical BEC experiment $\eta\sim 10-100$
\cite{SolenovMozyrskyRING,Boshier,Rb85exp}. If we keep the
tunneling exponent in (\ref{eq:gap}) equals to $\sim 1$ we obtain
$\mathrm{SNR}\sim 300-600$.

To conclude, we have demonstrated that a cold atom Josephson-ring
system can be described by an effective single-particle
Schrodinger equation with an effective potential that can be
controlled by the system's rotation. We analyzed the statistics of
the TOF measurements that allow one to determine the system's
current, i.e., phase across the Josephson junction. We found that
fidelity of such measurements is limited by the ground state
fluctuations and is controlled by the Tonks parameter.

Finally we conjecture that similar limitations are likely to take
place in superconducting flux qubits. While numerous studies of
such systems have been carried out, including the derivation of
the effective Schrodinger equation, etc., generalization of the
results obtained in this paper to the superconducting case,
however, does not seem to be straightforward due to a different
nature of the ground state for a fermionic system. Thus we believe
that the extension of our results to superconductors is an
interesting direction for future work.

{\acknowledgements We thank M. G. Boshier, A. J. Leggett, I.
Martin, V. Privman and E. Timmermans for valuable discussions and
comments. The work is supported by the US DOE. }


\end{document}